\documentclass{article}
      
      \newcommand{\bpi}{\mbox{\boldmath $\pi$}}
       
       \newcommand{\beq}{\begin{equation}}
       \newcommand{\eeq}{\end{equation}}
        \newcommand{\bea}{\begin{eqnarray}}
        \newcommand{\eea}{\end{eqnarray}}
\begin{document}
   \begin{center} {\large\bf Dipole transitions and Stark effect  in the charge-dyon  system}\\[5mm]
   {\large Levon Mardoyan$^{1}$, Armen Nersessian$^{1,2}$, Hayk Sarkisyan$^{1,3}$, Vahagn Yeghikyan$^{1}$} \end{center}
     {\it ${}^{1}$  Yerevan State University, 1 Alex Manoogian St.,
      Yerevan, 375025,  Armenia\\
      ${}^{2}$ Artsakh State University, 3 Mkhitar Gosh St.,  Stepanakert, Nagorny Karabakh\\
      $\;$Yerevan Physics Institute, 2 Alikhanian Brothers St., Yerevan, 375036, Armenia\\
       ${}^{3}$ Russian-Armenian University, 123 Hovsep Emin St,   Yerevan, 375051, Armenia}

        \begin{abstract}
        We consider the dipole transitions and
        the linear and quadratic Stark  effect in the  MICZ-Kepler system interpreting as a
        charge-dyon system.
        We show, that while the linear Stark effect in the ground state
         is proportional to azimuth quantum number(and to the sign of monopole
         number), the quadratic Stark effect in the ground state is independent on the
signs of azimuth and monopole numbers.
      \end{abstract}
      \section{Introduction}  The integrable system
      MICZ-Kepler has been suggested  independently by Zwanziger \cite{Z}  and
      by McIntosh and Cisneros  \cite{mic}
is defined by  the following
      Hamiltonian\footnote{through the paper  we use the {\sl Coulomb units},
      where the roles of mass,
      length and time units plays, respectively $\mu,\quad {\hbar^2}/{\mu\gamma},
      \quad {\hbar^3}/{\mu\gamma^2}
      $\cite{ll} (here $\mu$ is the mass of the particle,
      and  $\gamma$ is a Coulomb coupling constant.  We restrict ourself with positive $\gamma$ corresponding
      to the attractive MC-Kepler system).
      Particulary, the energy unit is $\mu\gamma^2/\hbar^2$. }
      \begin{equation}
       {\cal H}_{\rm MIC}=\frac{\bpi^2}{2} +\frac{{s}^2}{2
      r^2}-\frac{1}{r},\quad {\rm where }\quad
      [\pi_i,\pi_j]=-s\frac{\varepsilon_{ijk}x_k}{r^3},\quad
      [\pi_i,x_j]=-\imath\delta_{ij}.\label{1} \end{equation}
      Its distinguished peculiarity is the closed similarity with Coulomb
      problem, which insists in the existence
   of a
      hidden symmetry given by the angular momentum operator ${\bf L}$
and by the analog of Runge-Lenz vector, which are defined by the
expressions
      \begin{equation}
{\bf L}={\bf r}\times \bpi +s\frac{{\bf r}}{r},\qquad
      {\bf
      I}=\frac{1}{2}\left[\bpi\times{\bf L}-{\bf
      L}\times \bpi\right] +\frac{{\bf
      r}}{r} .\label{2} \end{equation}
This hidden symmetry exists due to the existence, in  the
Hamiltonian, of the specific centrifugal term $s^2/2r^2$. The
necessity of incorporation  of this term, in the Hamiltonian
describing the  motion of
      the electrically charged  non-relativistic particle
       in the field of Dyrac-dyon (particle carried both electric and magnetic
       charges).
However,  it  was realized decades ago, that the for the
consistent consideration of the particle-monopole systems this
term should be taken into account. Probably,  first time it was
pointed out by Zwanziger \cite{Z} an Schwinger \cite{sc}. Notice,
also, that this term yields, when we treat to obtain the
MICZ-Kepler system, similar to the Coulomb system,  from the
four-dimensional oscillator \cite{qmic}. Let us mention, in this
respect, that
 the Schroedinger
equation for the MICZ-Kepler system is equivalent to the
Schroedinger equation for the system of two well-separated BPS
monopoles/dyons (which possesses the Coulomb symmetry)
\cite{gibbons}. The actual observable difference   of the
MICZ-Kepler system   from  the Coulomb problem insists in the
change of the range of the total angular momentum
      from $l=0,1,\ldots$ to $l=|s|, |s|+1,  \ldots$
       (where  the monopole number $s$ takes (half-)integer values), which leads to
 the  $(2|s|+1)$-fold degeneracy of the ground state with respect to the azimuth quantum number.
On the other hand, the lifting of the low
        bound of the total angular momentum (and degeneracy of the ground state)
       is  the general feature  of the  quantum mechanical systems with monopoles.
       Another general feature of the quantum mechanical systems with
       monopoles is the change of the selection rules for the
       dipole transitions.
       Namely, while in conventional quantum mechanical
(spherically symmetric) models the selection rules are given by
the expressions
$  m'=m, \;l'=l-1;\quad m'=m\pm 1\; ,l'=l\pm 1\;\quad m'=m\pm 1\;
,l'=l\mp 1\;,$
in the charge-dyon system another transitions are also possible
(see \cite{tomilchik} and the references in \cite{comment}): 
$m'=m, \;l'=l\;\quad m'=m\pm 1\; ,l'=l .$ 
The
specific effect of the choice  of the Coulomb potential is hidden
symmetry,
 essentially simplifying the analyses
      of the  system. For example, it makes possible  the separation of variables in few coordinate systems
      \cite{book}.

It seems, that   MICZ-Kepler system could be useful
       for the modification of the existing models of
       quantum dots related with Coulomb problem, providing them with degenerate ground state.
 Moreover, that the search of the MICZ-Kepler system
 (as well as of the other quantum-mechanical systems with
 monopoles) in the condensed matter seems to be even more
 motivated than  in high energy physics and quantum field theory.
 Indeed, monopoles (and dyons) remains to be hypothetic particle, though their existence
 is admitted in modern field theoretical models. While in  the
 condensed matter the particle-monopole configuration could be viewed as short-distance approximation of the
 interaction of the behaviour of charged particle in the  vicinity of the pole of magnet.
Notice, that some attempt of incorporation of the Dirac monopole
in the
      quantum dot models has been performed in \cite{kurochkin}, giving the
      satisfactory interpretation of the
       experimental data. While  the paper \cite{haik}, where
     MIC-Kepler (charge-dyon) system in the spherical quantum well has been considered.
 Naively, one could expect, that $n$-th energy level of the MIC-Kepler system should be
identical with the $(n+|s|)$-th energy level of the Coulomb problem.
However,  the linear Stark effect in the
 MICZ-Kepler system (interpreted as a charge-dyon system)
 completely removes
the degeneracy of the energy levels
       in the charge-dyon system on the azimuth quantum number, in contrast with
       hydrogen atom  \cite{mara}.
Thus, one can believe, that  other observable differences between
MICZ-Kepler and Coulomb systems could also arise due to
interaction with external fields.

In this paper we study other possible differences in the behaviour
of the MICZ-Kepler system
 and Coulomb system, which could affect in the condensed matter applications.

At first,  we consider, for the completeness,   the  dipole
transitions in the MICZ-Kepler system generating by the planar
monochromatic electromagnetic wave. Their difference from the
dipole transitions in the so-called ``dyogen atom" (describing by
the MICZ-Kepler Hamiltonian  minus $s^2/2r^2$ term)
\cite{tomilchik} is in the value of the unessential constant.

Then  we consider  the   Stark effect in the charge-dyon system:
in contrast with dipole transitions, the specific choice of
potential is  important in this consideration. Particularly, since
the MICZ-Kepler system admits separation of variables in parabolic
coordinates,  both linear and quadratic Stark effects could be
calculated without any efforts.   In addition to linear Stark
effect calculated in \cite{mara}, we calculate the quadratic one,
 well as specify them for the ground
state of the MICZ-Kepler system.
 We find, that the ground state possesses both linear and quadratic
 Stark effect. The ground energy correction due to linear Stark
  is proportional to azimuth quantum number and to the  sign of monopole number,
   while the  quadratic Stark effect is independent on the signs of monopole
and azimuth quantum numbers.

 \setcounter{equation}{0}
      \section{Dipole transitions}

Let us consider the dipole transitions in the  MICZ-Kepler system
 interacting with planar monochromatic
electromagnetic wave, which are  completely similar to the  ones
in the ``dyogen atom" \cite{tomilchik}.

The wave is defined by the vector potential
\beq
{\bf A}=A_0 {\bf u}\cos ( \omega t-{\bf kr} )\;,\quad {\bf \nabla\cdot  A}=0
\eeq
where  ${\bf u}$ is polarization vector, ${\bf u k}=0$.
Assuming that the magnitude of this field is small enough, we could
represent the interaction energy as follows:

\beq
{\cal H}=\frac{(\bpi-{\bf A})^2}{2} +\frac{{s}^2}{2
      r^2}-\frac{1}{r}\approx {\cal H}_{\rm MIC}-{\bf A}{\bpi},
\eeq
where ${\cal H}_{\rm MIC}$ is defined by (\ref{1}).

For the calculation of the matrix element of dipole transitions we
shall use the  wavefunctions of the non-perturbed MICZ-Kepler
system in the spherical coordinates, which are the solutions of
the following spectral problem
 \beq  {\cal H}_{\rm MIC} \psi={\cal E}_{(0)}\psi\;,\qquad {\bf
       L}^2\psi=l(l+1)\psi\;,  \qquad L_3\psi= m\psi ,\label{sp} \eeq
These wavefunctions are given by the expressions
 \begin{equation} \psi_{nlm}({\bf r};{s})
      = c_{nl}r^l
      e^{-r/n}F(l-n+1, 2l+2,\frac{2r}{n}) d^l_{ms}(\theta)e^{im\varphi}.
      \label{wfs} \end{equation}
      Here
 $d^l_{ms}$ is the Wigner d-function, the energy spectrum is defined by the expression
 ${\cal E}_0=-{1}/{2n^2}$,  the quantum numbers $n, l, m$
  have the following ranges of definition
 \beq n=l+1, l+2,\ldots , \quad l=|s|,|s+1|,\ldots ,\quad m=-l,-l+1,\ldots,l-1,l.
\eeq
The normalization constant $c_{nl}$ is given by the expression (see, e.g. \cite{book})
\beq
c_{nl}=\frac{2^{l}}{ n^{l+2}(2l+1)!} \sqrt{\frac{(2l+1)(n+l)!}{\pi(n-l-1)!}}
\label{norm}\eeq
The matrix element for dipole transitions in the  long wave approximation
looks as follows \cite{davydov}
\label{c}\beq
M_{n,l,m|n',l',m'}=-\left[\frac{2\pi N}{V\omega}\right]^{1/2}{\bf u}\langle n,l,m|
\bpi|n',l',m'\rangle\;,
\eeq
where $N$ is the density of photons in the volume $V$.

Taking into account the commutation relation $[{\bf r},{\cal H}_{\rm MIC}]=\imath\bpi$,
one  can represent
the probability of transition from the state $(n,l,m)$ to the state $(n',l',m')$
in the unit time
in the following form
\beq
d{w}_{n,l,m|n',l',m'}=
\frac{N\omega^3}{2\pi}|{\bf u d }_{n,l,m|n',l',m'}|^2 d\Omega\;,\quad{\rm where}
\quad
{\bf d}_{n,l,m|n',l',m'}= ({\cal E}_{n',l',m'}-{\cal E}_{n,l,m})\langle n,l,m|{\bf r}| n',l',m' \rangle
\eeq
Straightforward calculation yields the result

$$
{\bf u d }_{n,l,m|n',l',m'}=I(n,l|n',l')
\left[ \frac{u_x+\imath u_y}{2}
\left( \frac{l+1}{2(2l+1)}\sqrt{(l+m)(l^2-s^2)}\delta_{m-1|m'}\delta_{l-1|l'}
\right.\right.$$

$$
\left. +\frac{\sqrt{(l+1)(l-m+1)(l-m+2)((l+1)^2-s^2)}}{2\sqrt{l+2}(l+1)(2l+2)}\delta_{m-1|m'}\delta_{l+1|l'}
+s\frac{\sqrt{(l-m+1)(l+m)}}{l(l+1)}
\delta_{m-1|m'}\delta_{l|l'} \right)
$$

$$
-\frac{u_x- \imath u_y}{2}
\left( \frac{l+2}{2(2l+3)}\sqrt{(l+m+2)((l+1)^2-s^2)}\delta_{m+1|m'}\delta_{l+1|l'}\right.
$$

$$
-\sqrt{\frac{l}{l+1}}\frac{\sqrt{(l-m-1)(l-m)(l^2-s^2)}}{l(2l-1)}\delta_{m+1|m'}\delta_{l-1|l'}
%
\left. +s\frac{\sqrt{(l+m+1)(l-m)}}{l(l+1)}
\delta_{m+1|m'}\delta_{l|l'} \right)+
$$
\beq
u_z\left( \frac{\sqrt{(l+1)(l^2-m^2)(l^2-s^2)}}{\sqrt{l}l(2l+1)}\delta_{l-1|l'}+\right.
\left.\left.\frac{\sqrt{(l+1)((l+1)^2-m^2)((l+1)^2-s^2)}}{\sqrt{l+2}(l+1)(2l+1)}\delta_{l+1|l'}
+s\frac{m}{l(l+1)}\delta_{l|l'}\right)\delta_{m|m'}\right]
\eeq
where
\beq
I(n,l|n',l')=\int_0^\infty c_{nl}c^*_{n'l'}r^{l+l'}{\rm e}^{-(\frac{r}{n'}+\frac{r}{n})}F(l-n+1,2l+2,\frac{2r}{n} )
F(l'-n'+1,2l'+2,\frac{2r}{n} ) r^3dr,
\eeq
and $c_{nm}$ is defined by (\ref{norm}).

It is seen  that the presence of Dirac monopole changes the selection rules of the system.
Namely, in the absence of monopole one has
\bea
&u_z\neq 0\;&:\qquad\qquad\qquad m=m', \;l'=l-1\;\\
&|u_x+\imath u_y|\neq 0 \;&:\quad m'=m\pm 1\; ,l'=l\pm 1\;; \qquad
m'=m\pm 1\; ,l'=l\mp 1\label{src} \eea In the presence of Dirac
monopole, when $s\neq 0$  another transitions are also possible
\cite{tomilchik,tomilchik1} \bea
&u_z\neq 0\;&: \; m=m', \;l'=l\;\\
&|u_x+\imath u_y|\neq 0 \;&:\; m'=m\pm 1\; ,l'=l\;\label{srm} \eea
 So,  the presence of monopole  makes the selection rules less rigorous. Namely,
besides (\ref{src}),  the transitions preserving the orbital quantum number
 $l$ become also allowed, (\ref{srm}). When the electromagnetic wave
 has transversal polarization ($u_z=0$),  the transitions
 preserving the orbital quantum number, and changing
 the azimuth quantum number become possible.
 When longitudal mode in the electromagnetic wave
 appears ($u_z\neq 0$), the transitions, preserving both orbital and azimuth
 quantum numbers are also admissible.

       \setcounter{equation}{0} \section{The Stark effect}
      The Hamiltonian of the MICZ-Kepler system (interpreting as charge-dyon system) in the
      external constant uniform  electric field is of the
      form: \begin{eqnarray} {\cal H}_{\rm Stark}={\cal H}_{MIC}+{{\bf E}{\bf r}}. \label{8}
      \end{eqnarray}
      Similar to the Coulomb system in the constant uniform magnetic field,
       this system possesses two
      constants of motion:
\beq
{J}\equiv {\bf n}_{E} {\bf L},\quad I= {\bf n}_E {\bf I}+
\frac{|{\bf E}|}{2}
({\bf n}_E\times {\bf r})^2,
\label{cm}\eeq
where ${\bf n}_E={\bf E}/|{\bf E}|$ is the unit vector directed along external electric field,
and ${\bf L}$ and ${\bf I}$ are given by the expressions (\ref{2}).
While the origin of the first constant of motion is obvious,
the validity of the second expression can
be checked by the straightforward calculation.
Due to the existence of the second constant
of motion the charge-dyon system interacting with the
external electric field
admits the separation of variables in parabolic coordinates. As a consequence,
completely  similar to the hydrogen atom,
one can  calculate the quadratic Stark effect in the charge-dyon system  \cite{ll}.
We assume that the electric field ${\bf E}$ is directed along
      positive $x_3$-semiaxes, and the force acting the electron is
       directed along negative  $x_3$-semiaxes.
We represent the momentum operator $\bpi$ as follows:
      \beq \bpi=-\imath {\bf\nabla }- {s}{\bf A}_{D}\;,\quad {\bf A}_{D}
      = \frac{1}{r(r - x_3)}\left(x_2, -x_1, 0\right)
      \eeq
      where ${\bf A}_{D}$ is the potential of the Dirac monopole with the
       singularity line directed along the positive semiaxis
      $x_{3}$.
      Choosing the parabolic coordinates
 $\xi,\eta \in [0, \infty), \, \varphi\in [0,
      2\pi)$   defined by the formulae \beq x_1+ix_2 = \sqrt{\xi
      \eta}e^{i\varphi}, \qquad x_3 = \frac{1}{2}(\xi - \eta).
      \label{parabolic}\eeq
   and making the substitution \beq
      \psi(\xi,\eta,\varphi) =
      \Phi_1(\xi) \Phi_2(\eta)\,\frac{e^{im\varphi}}{\sqrt{2\pi}}. \eeq
      we separate
       the variables in the Schr\"{o}dinger equation
 for the Hamiltonian (\ref{8}), and arrive at the system \cite{book}
 \begin{eqnarray}
 &&\frac{d}{d \xi}\left(\xi \frac{d\Phi_1}{d
      \xi}\right) + \left[\frac{\cal E}{2}\xi -\frac{|{\bf E}|}{4}\xi^2-
      \frac{(m+s)^2}{4\xi}
    \right]\Phi_1 = -\beta_1
     \Phi_1,
       \nonumber\\
       &&\frac{d}{d \eta}\left(\eta \frac{d\Phi_2}{d
      \eta}\right) + \left[\frac{ \cal E}{2}\eta +\frac{|{\bf E}|}{4}\eta^2 -
      \frac{(m-s)^2}{4\eta}\right]\Phi_2 =-\beta_2
    \Phi_2,\quad \beta_1+\beta_2=1. \end{eqnarray}
    It is seen, that $\beta_1-\beta_2$ is the eigenvalue of the operator $I$ in (\ref{cm}).

       For $s=0$ these
      equations coincide with the similar equations for the hydrogen atom in the
      parabolic coordinates \cite{ll}.
       Hence, similar to that, we can  consider the energy ${\cal E}$
      as a fixed parameter, and $\beta_{1,2}$ as the eigenvalues of corresponding operators.
      These quantities are defined after solving the above equations, as the functions
      on ${\cal E}$ and ${\bf E}$.
       Then, due to the relation $\beta_1+\beta_2=1$,   the energy ${\cal E}$
       becomes  a function
       on  the external field ${\bf E}$.
       Let us consider the terms containing the electric field $|{\bf E}|$
        as a perturbation.
       Thus, in zero approximation (${\bf E}=0$) we get
       \beq \Phi_1=\sqrt{\kappa}\Phi_{n_1,m+s}(\sqrt{\kappa}\xi ),\quad
       \Phi_1=\sqrt{\kappa}\Phi_{n_2,m-s}(\sqrt{\kappa}\eta ).
       \eeq
       Here
\beq \Phi_{p q}(x) =
      \frac{1}{|q|!} \sqrt{\frac{(p+|q|)!}{p!}} e^{-x/2}\left(x\right)^{|q|/2} {_1}F_1\left(-p;|q|+1;
      x\right). \eeq
      and
      $n_1$, $n_2$ are non-negative
      integers \beq
      \beta^{(0)}_1  = (n_1+ \frac{|m_1|+1}{2})\kappa ,\qquad
\beta^{0}_2  = (n_2+ \frac{|m_2|+1}{2})\kappa ,
     \eeq
     and
      \beq
      \kappa=\sqrt{-2 {\cal E}}, \quad m_a= m-(-1)^as\quad  a=1,2.\eeq
It is seen from the above expressions, that the calculation of the
 first  and second order corrections to the $\beta^{(0)}_{1,2}$ will be  completely
 similar to the ones in the Coulomb problem \cite{ll}, if one replaces $|m|\;\to \;|m+s|$ in $\beta_1$, $\Phi_1$,
 and $|m|\;\to \;|m-s|$ in $\beta_2$, $\Phi_2$.
 These substitutions yields the following expressions
      \beq
      \beta^{(1)}_a
=-\frac{(-1)^a |{\bf E}|}{4\kappa^2}
\left(6n_a^2+6n_a|m_a|+  m_a^2+6n_a+3 |m_a|  +2   \right)
\eeq
\beq
 \beta^{(2)}_a=-\frac{|{\bf E}|^2}{16\kappa^5}\left( |m_a|+2n_a+1\right)
 \left(     4m^2_a+17(2|m_a|n_a+2n^2_a+|m_a|+2n_a)+18\right).
\eeq
   Then
      we get
\beq
\beta^{0}_1+\beta^{(0)}_2=\kappa n,\quad
\beta^{(1)}_1+\beta^{(1)}_2=\frac{3|{\bf E}|}{2\kappa^2}A
,\quad
\beta^{(2)}_1+\beta^{(2)}_2=-\frac{|{\bf E}|^2}{16\kappa^5}B,
\eeq
where we introduce the notations
\beq A\equiv
 nn_- -\frac{ms}{3},\quad
 B\equiv17n^3-3n{n}^2_-+54An_-+19n-9n(m^2+s^2),
\eeq
and the quantum numbers
\beq
 n=n_1+n_2+\frac{|m+s|+|m-s|}{2}+1\;,
 \quad n_-\equiv n_1-n_2+\frac{|m+s|-|m-s|}{2}.\label{nn-}\eeq
Taking into account, that $\beta_1+\beta_2=1$, we get
\beq \kappa
n+\frac{3 |{\bf E}| A}{2\kappa^2}-\frac{|{\bf E}|^2 B}{16\kappa^5}=1 \eeq
     Iteratively solving this equation, we get
\beq
\kappa=\kappa_0+|{\bf E}|\kappa_1+|{\bf E}|^2\kappa_2\;,\qquad
\kappa_0=\frac{1}{n},\quad \kappa_1=-\frac{3An}{2},\quad \kappa_2=n^3\left(\frac{Bn}{16}-
\frac{9A^2}{2}\right).
\eeq
 Then, from  $E=-\kappa^2/2$   we find the energy of the system
      \beq
      {\cal E}=-\frac{1}{2n^2}+ \frac{3|{\bf E}|}{2}
      \left(nn_- -\frac{ms}{3}\right)
      -\frac{|{\bf E}|^2n^2}{16}\left(17n^4 -3(nn_- -3ms)^2-9n^2m^2+19n^2-9n^2s^2+21(ms)^2
      \right)
     \label{2stark} \eeq
   One can represent  the quantum numbers (\ref{nn-}) as follows
   \beq
   n=\left\{\begin{array}{ccc}
n_1+n_2+|s|+1&{\rm for }&|m|\leq |s|\\
n_1+n_2+|m|+1&{\rm for }&|m|>|s|
   \end{array}
 \right.\;, \qquad
 n_-=\left\{\begin{array}{ccc}
n_1-n_2+m\;{\rm sgn}\; s&{\rm for }&|m|\leq |s|\\
n_1-n_2+s\;{\rm sgn}\; m&{\rm for }&|m|>|s|
   \end{array}\right. .
 \eeq
The ground state of the non-perturbed charge-dyon system corresponds to the following
values of quantum numbers: $n_1=n_2=0$, $|m|\leq |s|$. Hence,
\beq
n=|s|+1,\quad n_-=m\;{\rm sgn}\; s\;,\quad  m=-|s|,-|s|+1,\ldots, |s|-1, |s|.
\eeq
Substituting these expressions in (\ref{2stark}), we get
\beq
{\cal E}_{0}=-\frac{1}{2(|s|+1)^2}+ m\; {\rm sgn}\; s \;|{\bf E}| \left(|s|+\frac{3}{2}\right)
-\frac{|{\bf E}|^2(|s|+1)^2}{16}\left[17(|s|+1)^4 +(|s|+1)^2(19-9s^2)- 6m^2\left(|s|+2\right)
   \right].
\eeq It is seen, that the ground state of the non-perturbed
charge-dyon system has $(2|s|+1)$-fold  degeneracy (by azimuth
quantum number $m$), while the linear Stark effect completely
removes the degeneracy on $m$. It is proportional to the azimuth
quantum number $m$, while its sign
 depends on the relative sign of monopole
number $s$ and $m$ (the linear Stark effect in the ``dyogen atom"
possesses similar properties \cite{tomilchik1}). In contrast to
linear Stark effect, the quadratic Stark effect of the ground
state   is independent neither on ${\rm sign }\; s$, no on ${\rm
sign }\; m$.

       \setcounter{equation}{0} \section{Conclusion}
We have studied the behavior of the MICZ-Kepler system,
 interpreted as a charge-dyon system, in the external fields,
  and for the testing its differences from the hydrogen atom.

We considered the dipole transitions in the MICZ-Kepler
interacting with planar monochromatic electromagnetic wave.
Similar to the other spherically-symmetric systems with monopole,
the MICZ-Kepler system admits dipole transitions which are
forbidden in conventional quantum mechanics. Namely, in
conventional quantum mechanics the dipole transitions should
satisfy the selection rules (\ref{src}). While in the systems with
monopolethe transitions satisfying the selection rules (\ref{srm})
are also possible. The behaviour of the MICZ-Kepler system in the
constant electric field is also different from the one in the
hydrogen atom. In the earlier work it was observed, that the
linear Stark effect completely removes the degeneracy of the
spectrum  in  the MICZ-Kepler system \cite{mara}. In the present
paper we calculated the quadratic Stark effect too. Considering
the ground state of the MICZ-Kepler charge-dyon system (which has
$(2|s|+1)$-fold degeneracy), we found that the  linear Stark
effect is proportional to the azimuth quantum number, and to the
sign of monopole number as well. While the quadratic Stark effect
depends on the absolute values of the azimuth and monopole
numbers.

Let us notice, that besides the system of two well-separated BPS
dyons \cite{gibbons}, another   integrable generalizations of the
MICZ-Kepler system on the curved spaces are also exist
\cite{micsphere}. In the context of quantum dots application these
systems could be viewed as a models with position-dependent
effective mass. Clearly, the dipole transitions in these systems
would be similar to those in conventional charge-dyon systems, but
the Stark effect could be essentially different. We are planning
to consider it elsewhere. Also, we expect, from some preliminary
analyses, that similar consideration could be performed after
incorporations the monopole in the more complicated integrable
system considered in \cite{rashid}. This system includes, besides
the Coulomb potential,
 also the oscillator one and the constant magnetic field.
Presented  study has been  performed for the testing the possible
consequences of incorporation of the   MICZ-Kepler system in  the
models of quantum dot. From this viewpoint,  MICZ-Kepler-like
generalization
 of the system \cite{rashid} seems to be especially interesting.

\vspace{5mm}

{\large Acknowledgments.} We   thank the authors of \cite{comment}
for the Comment, and Peter Horvathy and Rashid Nazmitdinov for the
useful remarks. Special thanks to
 Vadim Ohanyan for careful reading the manuscript  for the interest in work.
 The work was supported in part   by the grants INTAS-05-7928 and
NFSAT-CRDF  ARPI-3228-YE-04 and by the Armenian National Grant
 ``Semiconductor nanoelectronics".


\begin{thebibliography}{99}
        \bibitem{Z}D.~Zwanziger, 
Phys.\ Rev., {\bf 176}, 1480 (1968).

 \bibitem{mic}
H. ~McIntosh, A. ~Cisneros. 
 J. Math. Phys., {\bf 11}, 896 (1970).
 \bibitem{sc}J.Schwinger, Science, {\bf 165}, 757 (1969)
 \bibitem{ll} L.D.~Landau, E.M.~Lifshitz,{\sl Quantum Mechanics}, Oxford: Pergamon Press, 1977.
       \bibitem{qmic}
I.M. Mladenov and V.V. Tsanov, 
 J.\ Phys.\ A
 {\bf 20}, 5865 (1987);
J.\ Phys.\ A {\bf 32}, 3779 (1999).

 A.~Nersessian and
V.~Ter-Antonian,
  Mod.\ Phys.\ Lett.,\  A {\bf 9},  2431 (1994)
  Mod.\ Phys.\ Lett.\  A {\bf 10},  2633 (1995)

\bibitem{gibbons} G.~W.~Gibbons and N.~S.~Manton,
  Nucl.\ Phys.,\  B {\bf 274}, 183 (1986) .

  L.~G.~Feher and P.~A.~Horvathy,
  Phys.\ Lett. \ B  {\bf 183}, 182 (1987)
  [Erratum-ibid.\  {\bf 188B} (1987) 512].

 B.~Cordani, L.~G.~Feher and P.~A.~Horvathy,
  J.\ Math.\ Phys.,\  {\bf 31},  202, (1990).

 L.~G.~Feher,   J.\ Phys., \ {\bf A19}, 1259 (1986)
\bibitem{tomilchik}
 E.~A.~Tolkachev, L.~M.~Tomilchik and Y.~M.~Shnir,
  Yad.\ Fiz.\  {\bf 38}, 541 (1983);
  J.\ Phys.\ G {\bf 14},  1, (1988) 1.
  \bibitem{comment}
  E.~A.~Tolkachev and L.~M.~Tomilchik,
  arXiv:cond-mat/0610213.
 \bibitem{book}
 L.~G.~Mardoyan, A.~N.~Sissakian and V.M.~Ter-Antonyan.
      Int.\ J.\ Mod.\  Phys.\ A {\bf 12}, 237 (1997)

 L.~G.~Mardoyan, G.~S.~Pogosyan, A.~N.~Sissakian and V.~M.~ Ter-Antonyan,
 {\sl Quantum systems with hidden symmetry}, MAIK Publ., Moscow, 2006




 \bibitem{kurochkin} V.~V.~Gritsev and Yu.~A.~Kurochkin,  Phys.\ Rev. \ B {\bf
 64}, 035308 (2001).
  \bibitem{haik}
L.G. Mardoyan, L.S. Petrosyan, H.A. Sarkisyan, 
Phys.\ Rev., \ A {\bf 68}, 014103 (2003).

\bibitem{mara}
  L.~Mardoyan, A.~Nersessian and M.~Petrosyan,
  Theor.\ Math.\ Phys.,\  {\bf 140},  958 (2004)

\bibitem{davydov} A.~S.~Davydov, {\sl Quantum mechanics},  Nauka  Publ., Moscow, 1973
\bibitem{tomilchik1}
  E.~A.~Tolkachev, L.~M.~Tomilchik and Y.~M.~Shnir,
  Sov.\ J.\ Nucl.\ Phys.\  {\bf 50}, 275,  (1989);
  Sov.\ J.\ Nucl.\ Phys.\  {\bf 52},  916 (1990);

  Y.~M.~Shnir, E.~A.~Tolkachev and L.~M.~Tomilchik,
  Int.\ J.\ Mod.\ Phys.\ A {\bf 7},  3747 (1992).




 \bibitem{micsphere} V.~V.~Gritsev, Yu.~A.~Kurochkin, V.S.~Otchik, J.\ Phys. \ A {\bf 33}(2000), 4903.

  A.~Nersessian and G.~Pogosyan,
  Phys.\ Rev.\ A {\bf 63}, 020103(R) (2001)


  S.~Bellucci, A.~Nersessian and A.~Yeranyan,
  Phys.\ Rev.\ D {\bf 70}, 045006 (2004)

  G.~W.~Gibbons and C.~M.~Warnick,
  arXiv:hep-th/0609051.
\bibitem{rashid}N.~S.~Simonovi\'c and R.~G.~Nazmitdinov,
  Phys.\ Rev.\ B {\bf 67}, 041305(R) (2003)
      \end{thebibliography}
       \end{document}